\documentclass[reprint,superscriptaddress,aps,pra]{revtex4-1}
\usepackage{amssymb,amsmath}
\usepackage{mathrsfs}
\usepackage{braket}
\usepackage{color}
\usepackage{graphicx}% Include figure files
\usepackage{dcolumn}% Align table columns on decimal point
\usepackage{bm}% bold math

\usepackage[hypertexnames=false,colorlinks=true,linkcolor=red,citecolor=blue,urlcolor=blue]{hyperref} 
\usepackage{natbib}

\newcommand{\bk}{\bm{k}}
\newcommand{\br}{{r}}

\begin{document}

\preprint{APS/123-QED}

\title{Generation of magnonic squeezed state and its superposition in a hybrid qubit-magnon system}

\author{Gang Liu}
\email{gangliuphys@gmail.com}
%\thanks{\textcolor{blue}{\href{mailto:gangliuphys@gmail.com}{gangliuphys@gmail.com}} \\ 	  These authors contributed equally to this work.}
\affiliation{School of Physics, Zhejiang University, Hangzhou, China
}

\author{Junpeng Liu}
%\thanks{These authors contributed equally to this work.}
\affiliation{Lanzhou Center for Theoretical Physics, Key Laboratory of Theoretical Physics of Gansu Province, Key Laboratory
	of Quantum Theory and Applications of MoE, Gansu Provincial Research Center for Basic Disciplines of Quantum Physics,
	Lanzhou University, Lanzhou 730000, China}
\affiliation{Key Laboratory for Electronic Materials, College of Electrical Engineering, Northwest Minzu University,
	Lanzhou, Gansu 730000, China}

\author{Feng Qiao}
\email{qiaofeng1427@gmail.com}
\affiliation{School of Physical Science and Technology, Lanzhou University, Lanzhou 730030, China }

\author{Rong-Can Yang}
\email{rcyang@fjnu.edu.cn}
\affiliation{Fujian Provincial Key Laboratory of Quantum Manipulation and New Energy Materials, and College of Physics and Energy, Fujian Normal University, Fuzhou 350117, China}

\date{\today}

\begin{abstract}
	We propose a protocol for generating magnonic squeezed states and their superpositions in a hybrid system comprising a superconducting flux qubit magnetically coupled to the Kittel mode of a yttrium iron garnet sphere. The flux qubit provides an intrinsic longitudinal interaction with the magnon mode, which, under resonant microwave driving, gives rise to an effective qubit-state-dependent squeezing Hamiltonian. Numerical simulations incorporating realistic dissipation demonstrate that magnon quadrature noise reduction exceeding $8~\mathrm{dB}$ is achievable with experimentally accessible parameters.~By preparing the qubit in a superposition state followed by projective measurement, we further obtain symmetric and antisymmetric superpositions of orthogonally squeezed magnon states exhibiting clear phase-space interference fringes.~We discuss how the fourfold rotational symmetry of these states supports a bosonic logical encoding with potential for protecting against dominant error channels in magnonic platforms.
\end{abstract}

\maketitle

\section{\label{sec:intro}Introduction}

Squeezed states of harmonic oscillators, in which quantum fluctuations in one quadrature are reduced below the standard quantum limit, constitute a fundamental resource in quantum optics and continuous-variable quantum information processing.~They underpin quantum-enhanced sensing and precision measurement~\cite{schnabel2017squeezed,marino2008squeezed}, continuous-variable communication protocols~\cite{andersen2015hybrid}, and bosonic quantum error correction~\cite{gottesman2001encoding,albert2018performance}.~Of particular interest are superpositions of squeezed coherent states with distinct phase-space orientations, which combine quadrature squeezing with macroscopic quantum interference and serve as versatile resources for continuous-variable quantum computation~\cite{albarelli2018nongauss,albert2018performance}.~Realizing such states in microwave-frequency solid-state platforms compatible with superconducting circuits is therefore of considerable interest.

Hybrid quantum systems based on magnonics provide a natural setting for pursuing this direction~\cite{lachance2019hybrid,yuan2022quantum,Rameshti-2022,zuo2024cavity}.~In cavity magnonics, coherent interactions between microwave photons and collective spin excitations (magnons) in magnetic materials such as yttrium iron garnet (YIG) have been demonstrated and developed rapidly~\cite{Huebl2013prl,Tabuchi2014prl,Zhang2014prl}.~Magnons offer several attractive properties for quantum state engineering, including high spin density, low intrinsic dissipation at cryogenic temperatures, and wide tunability via external magnetic fields~\cite{tabuchi2015coherent,lachance2017resolving,lachance2020entanglement}.~In particular, the Kittel mode in a YIG sphere, corresponding to the uniform ferromagnetic resonance, provides a long-lived bosonic mode in the microwave domain and is well suited for integration with superconducting quantum circuits.~A further advantage of magnonic systems lies in their ability to couple to diverse physical degrees of freedom.~Magnons interact with microwave cavity photons via magnetic dipole coupling~\cite{Tabuchi2014prl,Zhang2014prl,shen2025nc}, with optical fields through magneto-optical effects~\cite{hisatomi2016bidirectional,osada2016cavity,zhang2016optomagnonic,Haigh2016prl}, and with mechanical modes via magnetostrictive interactions~\cite{zhang2016cavity,li2018magnon,Potts2021prx,shen2022prl,he2026cat}.~They can also couple to superconducting qubits, either indirectly through a shared cavity mode~\cite{tabuchi2015coherent,lachance2017resolving,lachance2020entanglement,wolski2020dissipation,xu2023quantum} or directly through magnetic interactions~\cite{Kounalakis2022prl,hou2024robust}.~These connections make hybrid magnonic platforms promising for the transfer and processing of nonclassical states across distinct subsystems~\cite{	Patton2013pra,Patton2013epl,lachance2019hybrid,yuan2022quantum}.~Leveraging these capabilities, a variety of nonclassical magnon states have been theoretically proposed, in some cases, experimentally explored, including squeezed states~\cite{li2019squeezed,zhang2021squeezed,2023guoprasqueezed,qian2024squeezing,xia2025magnon,d2st-rr91}, entangled states~\cite{qi2022generation,ren2022long,Hu:24,Golkar_2024,Fan2024,Zuo_2025,FAN20251958}, and cat states~\cite{Kounalakis2022prl,he2023pramc,hou2024robust,PhysRevApplied.21.044018,he2024pracats,jr6g-dcnp,liu2025magnoncat,Kani2025cat}.

Among these configurations, direct magnetic coupling between flux qubits and YIG magnons  has particularly attractived because it enables strong, controllable interactions and access to qubit-induced nonlinearity~\cite{Kounalakis2022prl,hou2024robust}.~The two persistent-current states of a flux qubit generate magnetic fields of opposite polarity at the YIG position, giving rise to a longitudinal interaction.~Compared with transverse coupling, longitudinal coupling enables conditional displacement and, when combined with strong driving, supports effective qubit-state-dependent parametric processes without the need for additional modulation tones~\cite{didier2015fast,hou2024robust}.~Exploiting this longitudinal coupling, several theoretical proposals have addressed nonclassical state generation in hybrid magnon-qubit systems, encompassing magnon blockade~\cite{Jin2023prablockade,8jy6-fp5x}, magnon entanglement~\cite{kounalakis2023entangled}, and macroscopic cat states~\cite{Kounalakis2022prl,hou2024robust,he2024pracats}.~Nevertheless, the simultaneous generation of squeezing and coherent superposition of squeezed magnon states within a unified protocol has received comparatively little attention.
In related qubit-oscillator systems based on superconducting circuits, qubit-conditional squeezing and the generation of superpositions of orthogonally squeezed states have been studied theoretically~\cite{Ayyash2024pra2,Ayyash2024pra,Ayyash2025arxiv}, yet analogous protocols in hybrid magnon-qubit platforms remain largely unexplored.~Such squeezed superposition states are of considerable interest: they combine reduced quadrature noise with macroscopic quantum interference in phase space, and their structured symmetries make them natural candidates for bosonic logical encodings relevant to quantum error correction~\cite{gottesman2001encoding,albert2018performance,Ayyash2024pra}.

Here we propose and analyze a protocol for generating strong magnonic squeezed states (MSS) and their superpositions (SMSS) in a flux-qubit-YIG hybrid system.~With the qubit biased at its optimal point and driven by a resonant microwave tone, we derive an effective conditional squeezing Hamiltonian that implements qubit-state-dependent parametric amplification of the magnon mode.~Numerical simulations incorporating realistic dissipation demonstrate that squeezing exceeding $8~\mathrm{dB}$ is achievable with experimentally accessible parameters.~By further preparing the qubit in a superposition state and performing projective measurement, we obtain symmetric and antisymmetric superpositions of orthogonally squeezed magnon states exhibiting clear phase-space interference fringes, which naturally admit a logical-qubit encoding with fourfold rotational symmetry relevant for bosonic quantum error correction~\cite{HopePRR2026}, or as a useful resource for non-Gaussian quantum computation \cite{PhysRevA.98.052350}.

The remainder of this paper is organized as follows.~Section~\ref{sec:level2} introduces the hybrid system and derives the effective conditional squeezing Hamiltonian.~Section~\ref{sec:level3} analyzes the squeezing dynamics under dissipation.~Section~\ref{sec:level4} discusses the preparation of SMSS and their connection to bosonic encodings and quantum error correction.~Finally, Sec.~\ref{sec:level5} summarizes our conclusions.

\begin{figure}[!t]
	\centering
	\includegraphics[width=0.96\linewidth]{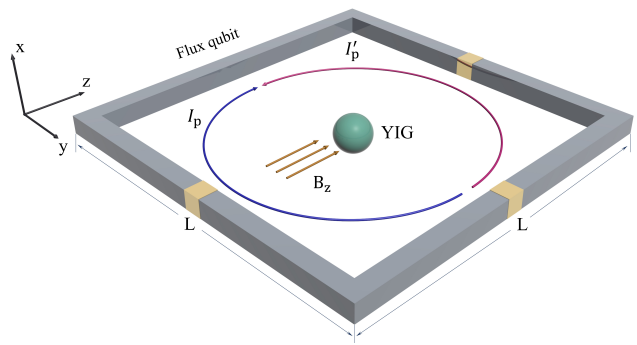}
	\caption{Schematic of the magnetically coupled hybrid system.~A YIG sphere is positioned at the center of a three-Josephson-junction superconducting flux qubit.~A static bias field $B_z$ is applied along the $z$ axis to tune the Kittel-mode frequency.~The two persistent-current states, carrying clockwise and counterclockwise currents $I_P$ and $I'_P$, generate opposite near fields at the sphere and thereby mediate a magnetic dipole (Zeeman) coupling between the qubit and the YIG.
	}
	\label{fig:fig1}
\end{figure}

\section{\label{sec:level2}Model and Hamiltonian}

We consider a superconducting flux qubit magnetically coupled to the uniform (Kittel) magnon mode of a YIG sphere placed in the near field of the qubit loop, as shown in Fig.~\ref{fig:fig1}.~In the persistent-current basis, the two circulating-current states generate near fields of opposite sign at the YIG position, giving rise to a state-dependent Zeeman interaction with the collective spin excitation.~The system Hamiltonian is (setting $\hbar=1$)
\begin{align}
	H = H_{\rm FQ}+H_m+H_{\rm int},
	\label{eq:H_decomp}
\end{align}
where $H_{\rm FQ}$ describes the flux qubit, $H_m$ the Kittel magnon mode, and $H_{\rm int}$ their magnetic interaction.

The flux qubit consists of three Josephson junctions with asymmetric Josephson energies, producing a double-well potential and two low-lying persistent-current states $\{\ket{\circlearrowright},\ket{\circlearrowleft}\}$ when biased near $\Phi_0/2$.~Projected onto this subspace, the effective two-level Hamiltonian reads
\begin{align}
	H_{\rm FQ}=-\frac{\epsilon_z}{2}\sigma_z-\frac{\epsilon_x}{2}\sigma_x ,
	\label{eq:HFQ_eps}
\end{align}
with Pauli operators $\sigma_{x,z}$ defined in the basis $\{\ket{\circlearrowright},\ket{\circlearrowleft}\}$ with $\sigma_x=\ket{\circlearrowright}\bra{\circlearrowleft}+\ket{\circlearrowleft}\bra{\circlearrowright}$ and $\sigma_z=\ket{\circlearrowright}\bra{\circlearrowright}-\ket{\circlearrowleft}\bra{\circlearrowleft}$.
The longitudinal bias
$\epsilon_z = 2I_p(\Phi_{\rm ext} - \Phi_0/2)$
is controlled by the external flux $\Phi_{\rm ext}$ threading the loop, where $I_p$ is the persistent current and $\Phi_0=h/2e$ is the flux quantum.~The transverse term $\epsilon_x$ is the tunneling amplitude between the two current states and can be tuned by replacing the smaller junction with a dc SQUID.

%The YIG sphere is modeled as a ferromagnetic spin ensemble subject to a static bias field $B_z\hat z$.~A minimal microscopic description is the isotropic Heisenberg Hamiltonian
%\begin{align}
%	H_{\rm spin}=-2J\sum_{\langle i,j\rangle}\mathbf{S}_i\!\cdot\!\mathbf{S}_j
%	+ g_e\mu_B B_z\sum_i S_i^z ,
%	\label{eq:Hspin_raw}
%\end{align}
%where $J>0$ is the nearest-neighbor exchange coupling, $\langle i,j\rangle$ denotes nearest neighbors, $g_e$ is the electron $g$ factor, and $\mu_B$ is the Bohr magneton.~Applying the Holstein-Primakoff transformation and working in the weak-excitation regime ($\langle m_i^\dagger m_i\rangle\ll 2S$) yields the leading-order relations
%\begin{align}
%	S_i^+ \simeq \sqrt{2S}\,m_i,\ 
%	S_i^- \simeq \sqrt{2S}\,m_i^\dagger,\ 
%	S_i^z = S - m_i^\dagger m_i .
%	\label{eq:HP_approx}
%\end{align}
%Diagonalization proceeds via plane-wave modes $m_{\bk}=N^{-1/2}\sum_i e^{-i\bk\cdot\br_i}m_i$, giving in the long-wavelength limit
%\begin{align}
%	H_m=\sum_{\bk}\omega(\bk)m_{\bk}^\dagger m_{\bk},
%	\label{eq:dispersion_lw}
%\end{align}
%where $\omega(\bk)\simeq g_e\mu_B B_z + 4SJ a_0^2|\bk|^2$ is the dispersion relation and $a_0$ is the lattice constant.~
In the present geometry, the spatially smooth component of the qubit-generated near field predominantly addresses the uniform precession mode; hence we retain only the Kittel mode ${\bk}=0$ and write
\begin{align}
	H_m=\omega_m\, m^\dagger m,
	\label{eq:Hm_kittel_m}
\end{align}
where $m$ ($m^\dagger$) annihilates (creates) a magnon, and $\omega_m = g_e\mu_B B_z$ is tunable via the bias magnetic field $B_z$.

\begin{figure}[!t]
	\centering
	\includegraphics[width=0.88\linewidth]{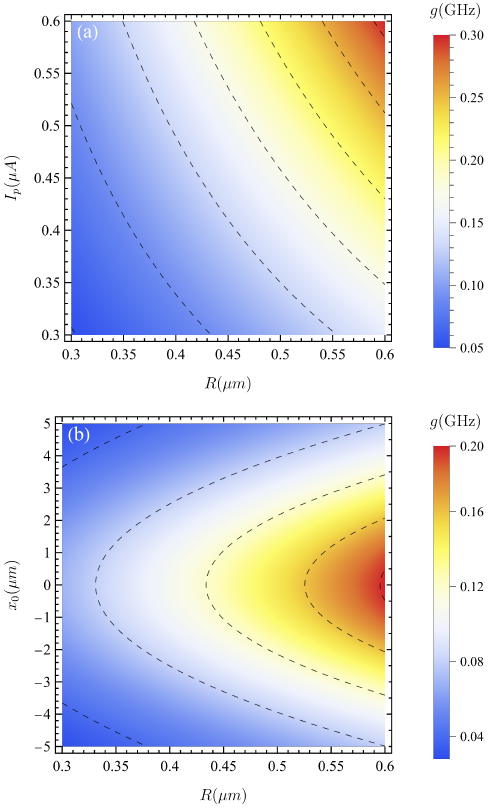}
	\caption{\label{fig:g_maps}
		(a) Magnon-qubit coupling strength $g$ as a function of the YIG radius $R$ and the flux-qubit persistent current $I_p$, with the loop side length fixed at $L=10~\mu{\rm m}$ and the sphere centered at the loop center $\br_0=\mathbf{0}$.
		The coupling is evaluated from Eq.~\eqref{eq:g_beff_compact1} using the Biot-Savart field of the square loop; in panel (a) we employ the point-sphere approximation $B_{\rm eff}^x\simeq B_{\rm loop}^x(\br_0)$ with $N=\rho(4\pi R^3/3)$.
		Dashed curves show isovalue-line $g$ contours from $0.05$ to $0.29$ in steps of $0.04$.
		(b) Effective coupling $g(\br_0,R)$ versus the radius $R$ and the out-of-plane offset $x_0$ for a fixed $I_p=0.4~\mu{\rm A}$.
		The square loop lies in the $y$-$z$ plane (normal along $\hat x$), and the sphere center is chosen as $\br_0=(x_0,0,0)$ along the symmetry axis.
		In panel (b), $B_{\rm eff}^x$ is obtained from the volume average over the sphere according to Eq.~\eqref{eq:g_beff_compact2}.
		Dashed curves show sovalue-line $g$ contours from $0.04$ to $0.20$ in steps of $0.04$.
	}
	\label{fig:fig12}
\end{figure}

The persistent current of the flux qubit generates a quasistatic magnetic near field at the YIG position.~In the persistent-current basis, the two circulating-current states $\{\ket{\circlearrowright},\ket{\circlearrowleft}\}$ produce magnetic fields of opposite sign.~We therefore write the qubit-generated field as $\mathbf{B}_{\rm FQ}(\br)=\sigma_z  \mathbf{B}_{\rm loop}(\br)$, and $\mathbf{B}_{\rm loop}(\br)$ denotes the field amplitude produced at $\br$ for a fixed circulating-current direction.

In the coordinate system of Fig.~\ref{fig:fig1}, the square loop lies in the $y$-$z$ plane and $\hat x$ denotes the loop-normal direction.~Near the loop center (and for symmetric placements), the qubit near field is dominated by its normal component, so that we retain $B_{\rm loop}^x(\br)$ and write the Zeeman interaction between the qubit field and the YIG spins as
\begin{align}
	H_{\rm int}&= g_e\mu_B \sum_i \mathbf{B}_{\rm FQ}(\br_i)\cdot \mathbf{S}_i\nonumber\\
	&\simeq g_e\mu_B\,\sigma_z \sum_i B_{\rm loop}^x(\br_i)\,S_i^x ,
	\label{eq:Hint_discrete}
\end{align}
where $\br_i$ labels the $i$-th spin location.~Away from the symmetry point, the loop field generally acquires in-plane components; these contributions are smaller in the parameter regime considered here and are neglected for clarity.~If the field variation across the sphere is small, the interaction predominantly addresses the uniform (Kittel) mode, and one may reduce Eq.~\eqref{eq:Hint_discrete} to the standard longitudinal coupling,
\begin{align}
	H_{\rm int}= g(\br_0,R)\,\sigma_z\,(m+m^\dagger).
	\label{eq:Hint_longitudinal}
\end{align}
To account for the finite sphere size and its placement relative to the loop, we express the coupling strength in terms of a volume-averaged effective field,
\begin{subequations}\label{eq:g_beff_compact}
	\begin{align}
		g(\br_0,R) &= g_e\mu_B\, B_{\rm eff}^x(\br_0,R)\sqrt{\frac{NS}{2}},
		\label{eq:g_beff_compact1}\\
		B_{\rm eff}^x(\br_0,R) &= \frac{1}{V}\int_{V(\br_0,R)} d^3r\, B_{\rm loop}^x(\br),
		\label{eq:g_beff_compact2}
	\end{align}
\end{subequations}
where $\br_0$ and $R$ are the sphere center and radius, $V=4\pi R^3/3$, and $N=\rho V$ is the total number of spins (with spin density $\rho$ and spin quantum number $S$).~Equation~\eqref{eq:g_beff_compact} makes explicit the trade-off relevant to experiments: increasing $R$ enhances the collective factor $\sqrt{N}\propto R^{3/2}$, while nonuniformity of the near field can reduce $B_{\rm eff}^x(\br_0,R)$ when the sphere samples regions of weaker field and larger gradients.~In the point-sphere approximation, one may set $B_{\rm eff}^x(r_0,R)\simeq B_{\rm loop}^x(r_0)\equiv B_{\rm FQ}^x$, yielding $g\simeq g_e\mu_B B_{\rm FQ}^x\sqrt{NS/2}$.

Using representative parameters ($R=0.5~\mu{\rm m}$, $\rho=2.1\times 10^{22}~{\rm cm}^{-3}$, $L=10~\mu{\rm m}$, $I_p=0.4~\mu{\rm A}$), we estimate $N\simeq 1.1\times 10^{10}$.~Approximating the loop-center field as $B_{\rm FQ}^x\simeq 2\sqrt{2}\mu_0 I_p/(\pi L)$ gives $g\simeq 2\pi\times 0.15~{\rm GHz}$ for $S=5/2$.~To guide experimental design beyond the point-sphere approximation, we compute $B_{\rm eff}^x(r_0,R)$ from the Biot-Savart field of the loop and summarize the resulting coupling strength in Fig.~\ref{fig:g_maps}.~Figure~\ref{fig:g_maps}(a) shows $g(R,I_p)$ for a sphere centered at the loop center $r_0={0}$ (with fixed $L$), while Fig.~\ref{fig:g_maps}(b) shows $g(R,x_0)$ for fixed $I_p = 0.4~\mu{\rm A}$, directly quantifying the dependence on sphere size and out-of-plane placement $\br_0=(x_0,0,0)$.

Collecting Eqs.~\eqref{eq:HFQ_eps}, \eqref{eq:Hm_kittel_m}, and \eqref{eq:Hint_longitudinal} (with constant $g$), the hybrid flux-qubit-magnon system is governed by
\begin{align}
	H = \omega_m\, m^\dagger m - \frac{\epsilon_z}{2}\sigma_z - \frac{\epsilon_x}{2}\sigma_x
	+ g(m+m^\dagger)\sigma_z,
	\label{eq:Hori}
\end{align}
which features an intrinsic longitudinal coupling between the flux qubit and the Kittel magnon mode in the persistent-current basis.

To engineer the magnon squeezing state, we apply a microwave drive $H_d = \Omega\cos(\omega_p t + \phi)(\sigma_x - \sigma_z)/\sqrt{2}% = -\Omega\cos(\omega_p t)(\sigma_x - \sigma_z)/\sqrt{2}
$ to qubit,
where $\Omega$ ($\omega_p$) is the drive amplitude (frequency) and $\phi$ is the initial phase, which is fixed at $\pi$ for optimal squeezing.~To diagonalize the flux-qubit Hamiltonian, we introduce the dressed-state basis $\ket{g} = \cos(\theta/2)\ket{\circlearrowleft} + \sin(\theta/2)\ket{\circlearrowright}$ and $\ket{e} = \sin(\theta/2)\ket{\circlearrowleft} - \cos(\theta/2)\ket{\circlearrowright}$, where $\ket{g}$ and $\ket{e}$ denote the ground and excited states of the flux qubit, respectively.~The flux angle is defined as $\theta = \arctan(\varepsilon_x/\varepsilon_z)$, and the qubit energy splitting between the dressed states is $\nu = \sqrt{\varepsilon_z^2 + \varepsilon_x^2}$.~We work at the optimal point $\theta=\pi/4$, for which $(\sigma_x-\sigma_z)/\sqrt{2}=\bar{\sigma}_x$.~In this dressed basis, the total Hamiltonian transforms to
\begin{align}\label{eq:Htot}
	H_\text{tot} =&\ \omega_m m^\dagger m + \frac{\nu}{2}\bar{\sigma}_z + g_x (m + m^\dagger) \bar{\sigma}_x \nonumber\\ 
	&\  + g_z(m + m^\dagger)\bar{\sigma}_z - \Omega\cos(\omega_p t )\bar{\sigma}_x, 
\end{align}
where $g_z = g\cos\theta$ and $g_x = g\sin\theta$ represent the longitudinal and transverse coupling strengths, respectively.~Here, $\bar{\sigma}_z \equiv \ket{e}\bra{e} - \ket{g}\bra{g}$ is the Pauli operator in the dressed basis, while $\bar{\sigma}_+ = \ket{e}\bra{g}$ and $\bar{\sigma}_- = \ket{g}\bra{e}$ are the raising and lowering operators, and $\bar{\sigma}_x = \bar{\sigma}_+ + \bar{\sigma}_-$.~Note that both the transverse and longitudinal couplings can give rise to a nonlinear two-magnon exchange interaction between the qubit and the magnon mode in the YIG sphere.~These interactions constitute key ingredients for generating magnon squeezing.

Next, we perform a unitary transformation $U = \exp[- i( m^\dagger m + \bar{\sigma}_z )\omega_p t /2]$.~The transformed Hamiltonian becomes
\begin{align}
	H_\text{tot}^\prime =&\ \Delta_m m^\dagger m  + \frac{\Delta_\nu}{2}\bar{\sigma}_z  - \frac{\Omega}{2}\bar{\sigma}_x\nonumber\\
	& + g_x (m\bar{\sigma}_- e^{-\frac{3}{2}i\omega_p t} + m\bar{\sigma}_+ e^{\frac{1}{2}i\omega_p t} + \text{H.c.})\nonumber\\
	& + g_z (m e^{-\frac{1}{2}i\omega_p t} + m^\dagger e^{\frac{1}{2}i\omega_p t})\bar{\sigma}_z,
	\label{eq:Htotprime}
\end{align}
where $\Delta_m = \omega_m - \omega_p/2$ and $\Delta_\nu = \nu - \omega_p$.~Applying the rotating-wave approximation (RWA) with respect to the drive frequency $\omega_p$, we retain the resonant drive term $ {\Omega}/{2}\bar{\sigma}_x$ and discard components oscillating at $\pm 2\omega_p$.
%\textcolor{red}{Although these counter-rotating drive components do not contribute at first order, they generate a static Bloch-Siegert correction at second order, leading to a small renormalization of the effective qubit detuning.~In practice, this shift can be absorbed into $\Delta_m$ when comparing the full driven dynamics with the effective model.}
To proceed further, we use an effective-Hamiltonian treatment to capture the time-averaged dynamics of this highly detuned system.

We express the time-dependent component of Eq.~\eqref{eq:Htotprime} as $H_\text{tot}^\prime(t)$, which admits the decomposition
\begin{align}
	H_\text{tot}^\prime(t) = \sum_{m=1,2,3} \left(h^\dagger_m e^{i\delta_m t} + \text{H.c.}\right),
	\label{eq:Htotprimet}
\end{align}
where the constituent operators are defined as $h^\dagger_1 = g_x m\bar{\sigma}_+$, $h^\dagger_2 = g_x m^\dagger\bar{\sigma}_+$, and $h^\dagger_3 = g_z m^\dagger\bar{\sigma}_z$, with corresponding detunings $\delta_1 = \delta_3 = \omega_p/2$ and $\delta_2 = 3\omega_p/2$.~In the large-detuning regime $\omega_p \gg \{\Delta_m, \Delta\nu, g_x, g_z, \Omega\}$, the detunings $\delta_m$ are sufficiently large that we can substitute Eq.~\eqref{eq:Htotprimet} into the standard effective-Hamiltonian expression and obtain the effective Hamiltonian for $H_\text{tot}^\prime(t)$, denoted as $H_\text{tot}^{\prime\prime}(t)$, in the form~\cite{james2007effective}
\begin{align}
	H_\text{tot}^{\prime\prime}(t) =& \sum_{m,n=1}^3 \frac{-1}{\delta_n} \left[h^\dagger_m h^\dagger_n e^{i(\delta_m+\delta_n)t} + h_m h_n^\dagger e^{-i(\delta_m-\delta_n)t}\right.\nonumber\\
	&\left.~- h^\dagger_m h_n e^{i(\delta_m-\delta_n)t} - h_m h_n e^{-i(\delta_m+\delta_n)t}\right],
	\label{eq:Htotprimeprimet1}
\end{align}
Then, applying the RWA to neglect the rapidly oscillating terms, Eq.~\eqref{eq:Htotprimeprimet1} can be simplified to
\begin{align}\label{eq:Htotprimeprimet2}
	H_\text{tot}^{\prime\prime}(t) =& \sum_{m=1}^3 \frac{1}{\delta_m} [h^\dagger_m, h_m] \nonumber\\
	&+ \sum_{m,n=1,2,3}^{m<n} \frac{1}{\bar{\delta}_{mn}} \{ [h^\dagger_m, h_n] e^{i(\delta_m-\delta_n)t} + \text{H.c.}\},
\end{align}
with $\bar{\delta}_{mn} \equiv (\delta_m + \delta_n)/2$.~Substituting the operator commutators $[h_m, h_n]$ derived from Eq.~(\ref{eq:Htotprimet}) into Eq.~\eqref{eq:Htotprimeprimet2}, we obtain
\begin{align}\label{eq:Htotprimeprimet3}
	H_\text{tot}^{\prime\prime}(t) = &\frac{8g_x^2}{3\omega_p}\left(2m^\dagger m\ket{e}\bra{e} + \ket{e}\bra{e} - m^\dagger m\right)\nonumber\\
	& + \frac{g_x^2}{\omega_p}\left(m^2\bar{\sigma}_z e^{-i\omega_p t} + m^{\dagger 2}\bar{\sigma}_z e^{i\omega_p t}\right)\nonumber\\
	& - \frac{g_x g_z}{\omega_p}(1 + 2m^\dagger m)\left(\bar{\sigma}_+ e^{i\omega_p t} + \bar{\sigma}_- e^{-i\omega_p t}\right)\nonumber\\
	& - \frac{4g_x g_z}{\omega_p}\left(m^{\dagger 2}\bar{\sigma}_- + m^2\bar{\sigma}_+\right).    
\end{align}
The first line of Eq.~\eqref{eq:Htotprimeprimet3} contains static second-order shifts induced by the transverse coupling.~These terms renormalize the slow detunings and give rise to an ac-Stark shift.~In the presence of strong driving, rapidly oscillating and nonresonant contributions are averaged out under the RWA, so that their net effect can be absorbed into an effective detuning.
By neglecting the rapidly oscillating terms in Eq.~\eqref{eq:Htotprimeprimet3} and retaining the time-independent contributions together with those in Eq.~\eqref{eq:Htotprime}, we obtain the following effective Hamiltonian:
\begin{align}
	H_\text{eff} =&\ (\Delta_m - \frac{8g_x^2}{3\omega_p})m^\dagger m  + \frac{\Delta_\nu}{2}\bar{\sigma}_z\nonumber\\
	& + \frac{8g_x^2}{3\omega_p}\left(2m^\dagger m\ket{e}\bra{e} + \ket{e}\bra{e}\right)\nonumber\\
	& - \frac{4g_x g_z}{\omega_p}\left(m^{\dagger 2}\bar{\sigma}_- + m^2\bar{\sigma}_+\right) - \frac{\Omega}{2} \bar{\sigma}_x.
	\label{eq:Heff}
\end{align}

To engineer a standard single-mode squeezing interaction, we move to the interaction picture defined by the strong drive,
$U=\exp[i(\Omega/2)\bar{\sigma}_x t]$, so that the Hamiltonian in this driven frame is
$H_\text{eff}^\prime = U H_{\rm eff}U^\dagger - iU\dot U^\dagger$.
In the regime $\Omega \gg 4|g_x g_z|/\omega_p$, terms that do not commute with $\bar{\sigma}_x$ acquire rapidly oscillating prefactors and average out under the RWA.~Retaining only the time-independent contribution along $\bar{\sigma}_x$, we obtain the effective Hamiltonian in the drive interaction picture,
\begin{align}
	H_\text{eff}^\prime
	= \Delta_m\, m^\dagger m
	+ g_{\text{cs}}\left(m^2 + m^{\dagger 2}\right)\bar{\sigma}_x ,
	\label{eq:Heffp}
\end{align}
which describes a qubit-state-dependent degenerate parametric-amplification process for the magnon mode and thus generates conditional magnon squeezing.~
Here $g_{\text{cs}}\equiv -2g_x g_z/\omega_p$ is the effective conditional-squeezing strength.
In the numerical simulations, we use the same reduced parametric-amplifier model as in Eq.~\eqref{eq:Heffp}, but with the detuning renormalized as $\Delta_m\rightarrow \Delta_{\rm eff}$ to account for static second-order frequency shifts arising from the transverse coupling and from the counter-rotating components of the drive.
Finally, moving to the interaction picture with respect to $\Delta_{\rm eff} m^\dagger m$, we obtain
\begin{align}
	H_{\rm cs}^{(I)}
	= g_{\text{cs}}\left( m^2 e^{-i2\Delta_{\rm eff} t} + m^{\dagger 2} e^{i2\Delta_{\rm eff} t}\right)\bar{\sigma}_x,
	\label{eq:conditional_squeezing}
\end{align}
which reduces to a time-independent squeezing interaction at effective resonance $\Delta_{\rm eff}\to 0$.
%In the regime $\Omega \gg 4|g_x g_z|/\omega_p$, all rapidly oscillating terms proportional to $\bar{\sigma}_y$ and $\bar{\sigma}_z$ average out under the RWA, leaving only the static contribution along $\bar{\sigma}_x$. 
The corresponding time-evolution operator generated by Eq.~\eqref{eq:conditional_squeezing} is
\begin{align}
	U_\text{cs}^{(I)}(t) = \ket{+}\bra{+}S(\xi(t)) + \ket{-}\bra{-}S(-\xi(t)),
\end{align}
where $|\pm\rangle$ are $\bar{\sigma}_x$ eigenstates, $S(\xi)=\exp[(\xi^* m^2 - \xi m^{\dagger 2})/2]$ is the squeezing operator, and $\xi(t) = -g_{\rm cs}(e^{i2\Delta_{\rm eff} t} - 1)/(2\Delta_{\rm eff})$ is the squeezing parameter.~At resonance $\Delta_{\rm eff}\to 0$, one recovers $\xi(t)\to -ig_{\rm cs}t$, and the two qubit eigenstates generate squeezing along orthogonal quadratures in magnon phase space.~When the qubit is prepared in an eigenstate of $\bar{\sigma}_x$ with eigenvalue $\pm 1$, the magnon evolution operator is $U(t) = S(\pm \xi(t))$.~Building on these results, we can describe a protocol to generate symmetric and antisymmetric superpositions of orthogonally squeezed states, analogous to the generation of even and odd cat states via conditional displacement in qubit-oscillator systems with linear coupling~\cite{AgarwalPRL2003,liao2016generation}.

\section{\label{sec:level3}GENERATION OF MAGNON SQUEEZING}

In Sec.~\ref{sec:level2}, we analytically derived an effective conditional squeezing Hamiltonian that enables the generation of magnon squeezing via unitary evolution in the absence of dissipation, as described by the effective Hamiltonian $H_\text{cs}^{(I)}$ in Eq.~\eqref{eq:conditional_squeezing} under the RWA.~In this section, we numerically investigate the magnon squeezing dynamics by incorporating dissipation and using experimentally feasible parameters.
%
%Specifically, we calculate the magnon squeezing using the effective Hamiltonian in Eq.~\eqref{eq:conditional_squeezing}, and assess the results against those obtained from the original driven longitudinal-transverse Rabi Hamiltonian in Eq.~\eqref{eq:Htot}, which involves no approximations. This comparison allows us to assess the validity of the effective model and identify the parameter regime in which it remains accurate. We then examine the influence of dissipation and extend the analysis to non-Gaussian superpositions of squeezed states.
%When the qubit is prepared in an eigenstate of $\bar{\sigma}_x$ with eigenvalue $s=\pm1$, the effective Hamiltonian reduces to $H_{\rm cs}^{(s)} = s g_{\rm cs}\left(m^{\dagger 2}+m^2\right)$, which corresponds to a time-independent single-mode parametric amplification Hamiltonian. The resulting unitary evolution is given by $U(t)=\exp\!\left[-is g_{\rm cs} t\left(m^{\dagger 2}+m^2\right)\right] = S(\xi)$ with squeezing parameter $\xi=2is g_{\rm cs} t$. The two qubit eigenstates thus generate squeezing along orthogonal quadratures in magnon phase space. 
%
Magnon squeezing is characterized by a reduction in the variance of a generalized quadrature operator $X_\theta = \cos(\theta) X_1 + \sin(\theta) X_2$ below the vacuum noise level, where $X_1 = (m + m^\dagger)/\sqrt{2}$ and $X_2 = i(m^\dagger - m)/\sqrt{2}$ denote the amplitude and phase quadratures, respectively.~We quantify squeezing through the principal quadrature variance $\zeta_B^2 \equiv\min_{\theta} (\Delta X_\theta)^2$, i.e., the minimum of $(\Delta X_\theta)^2$ with respect to $\theta$.~Squeezing is present when $\zeta_B^2 < V_\text{vac}$, where $V_\text{vac} = 1/2$ is the vacuum fluctuation level~\cite{ma2011quantum}.~The minimization can be performed analytically, yielding
\begin{equation}
	\zeta_B^2
	= 1 + 2 (\langle m^\dagger m\rangle - |\langle m\rangle|^2 )
	- 2|\langle m^2\rangle - \langle m\rangle^2 |,
\end{equation}
which depends only on the first- and second-order moments of the magnon mode.~The time dependence $\zeta_B^2(t)$ directly captures the buildup of magnon squeezing and identifies the point of maximal noise reduction.

\begin{figure}[tp]
	\centering
	\includegraphics[width=0.86\linewidth]{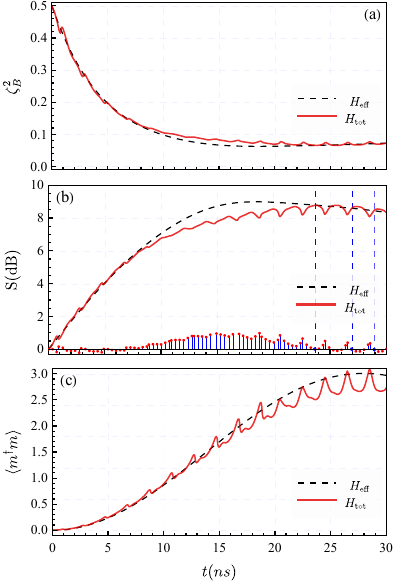}
	\caption{Time evolution of (a) the minimum quadrature variance $\zeta_B^2$, (b) the corresponding squeezing degree $S$, and (c) the mean magnon number $\langle m^{\dagger}m\rangle$.~The red solid curves correspond to the dynamics governed by the total Hamiltonian in Eq.~\eqref{eq:Htot} and obtained by solving the master equation~\eqref{eq:me}, whereas the black dashed curves are calculated from the effective Hamiltonian in Eq.~\eqref{eq:conditional_squeezing} by solving the corresponding effective master equation~\eqref{eq:me2}.~The system is initialized in the magnon vacuum state, while the qubit prepared in the $\bar{\sigma}x$ eigenstate $(\ket{g} + \ket{e})/\sqrt{2}$.~The parameters are chosen as $\omega/2\pi = 1.513~{\rm GHz}, \omega_p/2\pi = 3.002~{\rm GHz}, \Omega/2\pi = 0.5~{\rm GHz}, \phi=\pi$, $g=2\pi\times0.15~\mathrm{GHz}$, $g_x=g_z=g/\sqrt{2}$ with $\theta=\pi/4$, and $\nu=2\pi\times3~\mathrm{GHz}$.~The dissipation rates are set to $\kappa/2\pi = 0.5~\mathrm{MHz}$, $\gamma/2\pi = 3~\mathrm{kHz}$, and $\gamma_\phi = \gamma$, with a bath temperature $T=10~\text{mK}$.~}
	\label{fig:fig2}
\end{figure}

To analyze the robustness of conditional magnon squeezing under realistic experimental conditions, we include dissipation and decoherence by solving the Lindblad master equation.~The dynamics governed by the full driven magnon-qubit Hamiltonian $H_{\rm tot}$ in the laboratory frame are described by
\begin{align}\label{eq:me}
	\frac{d\rho}{dt} &= -i[H_{\text{tot}}, \rho]  + \frac{\kappa}{2}(\bar{n}_{{m}} + 1)\mathcal{L}[m]\rho + \frac{\kappa}{2}\bar{n}_{{m}}\mathcal{L}[m^\dagger]\rho  \nonumber\\
	& +  \frac{\gamma}{2}(\bar{n}_{{q}} + 1)\mathcal{L}[\sigma_-]\rho + \frac{\gamma}{2}\bar{n}_{{q}}\mathcal{L}[\sigma_+]\rho + \frac{\gamma_\phi}{4} \mathcal{L}[{\sigma}_z]\rho ,
\end{align}
where $\rho$ is the density operator of the system in the original frame, and $\mathcal{L}[o]\rho = 2o\rho o^\dagger - o^\dagger o \rho - \rho o^\dagger o$ denotes the standard Lindblad superoperator% for an operator $o$ (with $o = \{m, {\sigma}_-, {\sigma}_z\}$)
.~Here, $\kappa$ and $\gamma$ represent the energy relaxation rates of the magnon and qubit, respectively, while $\gamma_\phi$ denotes the pure dephasing rate of the qubit.~The mean thermal occupation number is given by $\bar{n}_{{m(q)}} = [\exp(\omega_{m(q)}/k_B T) - 1]^{-1}$, where $T$ is the bath temperature and $k_B$ is the Boltzmann constant.~After applying the rotating-frame transformation and the approximations introduced in Sec.~\ref{sec:level2}, the effective dynamics governed by the Hamiltonian $H_{\rm cs}$ are described by
\begin{align}\label{eq:me2}
	\frac{d\tilde{\rho}}{dt} =& -i[H_{\text{cs}}^{(I)}, \tilde{\rho}]  + \frac{\kappa}{2}(\bar{n}_{{m}} + 1)\mathcal{L}[m]\tilde{\rho} + \frac{\kappa}{2}\bar{n}_{{m}}\mathcal{L}[m^\dagger]\tilde{\rho}\nonumber\\
	& + \frac{\gamma}{8}(2\bar{n}_{{q}} + 1)\mathcal{L}[\bar\sigma_x]\tilde{\rho},
\end{align}
where $\tilde{\rho}$ is the density operator in the rotating frame.

\begin{figure}[!t]
	\centering
	\includegraphics[width=0.82\linewidth]{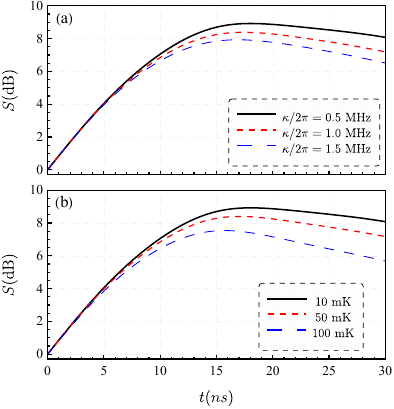}
	\caption{Time evolution of the squeezing degree $S$ (dB) for (a) several magnon damping rates $\kappa$ and (b) various bath temperatures $T$.~In panel (a) we fix $T=10~\mathrm{mK}$, while in panel (b) we set $\kappa/2\pi=0.5~\mathrm{MHz}$.~All other parameters are identical to those used in Fig.~\ref{fig:fig2}.}
	\label{fig:fig3}
\end{figure}

To validate the effective description, we numerically compute the principal quadrature variance $\zeta_B^2(t)$ of the magnon mode [Fig.~\ref{fig:fig2}(a)] from the \emph{post-selected} magnon state conditioned on the qubit measurement outcome $\bar{\sigma}_x = +1$, i.e., a measurement in the basis $(\ket{g} + \ket{e})/\sqrt{2}$.~We also compute the corresponding squeezing degree $S = -10 \log{10} [\zeta_B^2/V_\text{vac}]$ [Fig.~\ref{fig:fig2}(b)], in units of dB, and the mean magnon excitation [Fig.~\ref{fig:fig2}(c)], by solving the Lindblad master equations generated by the full Hamiltonian $H_{\rm tot}$ [Eq.~\eqref{eq:Htot}] (red solid) and by the effective conditional-squeezing Hamiltonian $H_{\rm cs}^{(I)}$ [Eq.~\eqref{eq:conditional_squeezing}] (black dashed), using identical initial conditions.~As shown in Fig.~\ref{fig:fig2}(b), both models predict peak magnon squeezing exceeding $8~\mathrm{dB}$ and agree closely over the full time evolution, indicating that $H_{\rm cs}^{(I)}$ provides an accurate description in this parameter regime.~The red markers at the bottom of Fig.~\ref{fig:fig2}(b) show the difference between the squeezing dynamics obtained from the effective and full Hamiltonians, while the blue markers indicate time points at which the discrepancy is smaller than $0.02~\mathrm{dB}$.~Figure~\ref{fig:fig2}(c) shows the corresponding evolution of the mean magnon occupation number.

To assess the impact of dissipation, we further analyze the time evolution of $S(t)$ in the presence of magnon damping and thermal noise.~Figure~\ref{fig:fig3}(a) illustrates the influence of the magnon dissipation rate on the squeezing.~Stronger damping clearly degrades the squeezing performance, with the most pronounced suppression occurring at early times when the maximum squeezing is reached.~For a relatively low damping rate, $\kappa/2\pi \approx 0.5~\mathrm{MHz}$, as reported in a recent experiment~\cite{shen2025nc}, the squeezing can still exceed $8~\mathrm{dB}$.~By contrast, at later evolution times the squeezing becomes much less sensitive to $\kappa$.~The influence of the bath temperature is shown in Fig.~\ref{fig:fig3}(b).~When the temperature is below a few tens of millikelvin, the thermal magnon occupation is negligible and the mode is initialized close to the vacuum state, so thermal fluctuations have little effect on the squeezing.~However, for temperatures above $\sim 100~\mathrm{mK}$, thermal excitations become significant and noticeably reduce the achievable squeezing.~Overall, dissipation limits the attainable squeezing depth and introduces an optimal evolution time, yet substantial quadrature-noise reduction below the vacuum level persists over experimentally relevant timescales.~In particular, for $\kappa \ll g_{\rm cs}$ and low thermal occupation, the maximum squeezing remains close to the ideal unitary prediction.~These results demonstrate that conditional magnon squeezing generated by $H_{\rm cs}^{(I)}$ is robust against realistic dissipation, providing a reliable platform for further non-Gaussian state engineering.~In the following section, we exploit this robustness to generate superpositions of squeezed states.

\begin{figure}[!t]
	\centering
	\includegraphics[width=0.86\linewidth]{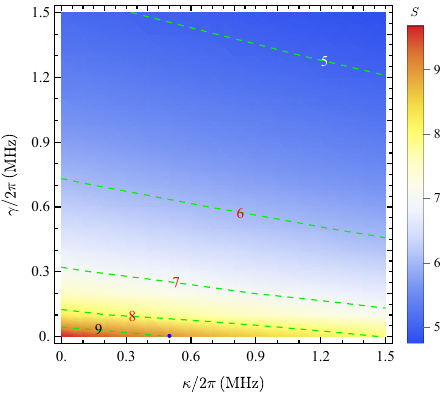}
	\caption{(a) Maximum squeezing $S$ as a function of the magnon and qubit dissipation rates $\kappa$ and $\gamma$.~For each pair $(\kappa,\gamma)$, the reported value corresponds to the largest squeezing obtained by optimizing the evolution time and the squeezing angle.~The blue dot marks the parameter set used in Fig.~\ref{fig:fig2}.~All other parameters are the same as those in Fig.~\ref{fig:fig2}.}
	\label{fig:fig4}
\end{figure}

In Fig.~\ref{fig:fig4}(a), we plot the maximum magnon squeezing as a function of the magnon and qubit dissipation rates.~Within the RWA used in Eq.~\eqref{eq:me2}, the contribution from qubit pure dephasing is negligible, so the squeezing is primarily limited by energy relaxation.~Both dissipation channels noticeably suppress the attainable squeezing.~In practice, however, qubit relaxation rates achieved in experiments~\cite{whh2019cat,Ren2022} are typically far smaller than the magnon damping of a YIG sphere, which is often on the order of $\kappa/2\pi \sim 1~\mathrm{MHz}$ due to intrinsic losses~\cite{Tabuchi2014prl,Zhang2014prl}.~Consequently, the relatively large magnon dissipation is the main bottleneck preventing substantially stronger squeezing in our scheme.

\begin{figure}[!t]
	\centering
	\includegraphics[width=0.98\linewidth]{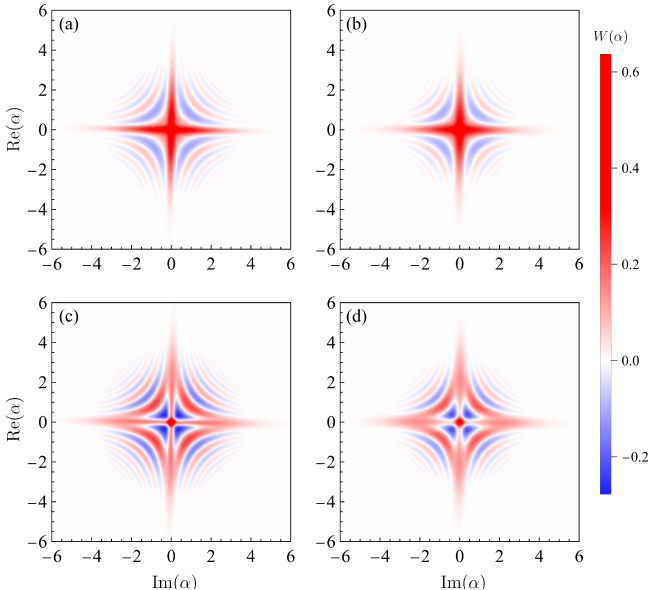}
	\caption{Wigner functions $W(\alpha)$ of the (a,b) symmetric and (c,d) antisymmetric superpositions of orthogonally squeezed magnon states.~The states are generated via a conditional squeezing protocol governed by $H_{\rm cs}^{(I)}$.~Starting from an initial state $\ket{0}\otimes(\ket{+_x}+\ket{-_x})/\sqrt{2}$, the system evolves into an entangled qubit-magnon state, followed by a projective measurement of the qubit in the $\bar{\sigma}_x$ basis.~Depending on the measurement outcome, the magnon is projected onto the even or odd superposition of squeezed vacuum states.~The dissipation-free evolution is depicted in panels (a) and (c).~In contrast, panels (b) and (d) account for environmental decoherence with: $\kappa/2\pi = 0.5\ \text{MHz}$, $\gamma/2\pi = 3\ \text{kHz}$, and $T = 10\ \text{mK}$.~Here an evolution time is chosen as $t = 29~{ns}$, as this yields a squeezing parameter whose real and imaginary parts have equal magnitudes, a condition crucial for the subsequent state generation.
	}
	\label{fig:fig5}
\end{figure}

\section{\label{sec:level4}Superpositions of squeezed magnon states %and application to quantum codes
}

Building on the Hamiltonian $H_{\rm cs}^{(I)}$, which generates orthogonally squeezed magnon states conditioned on the qubit state, %when the qubit is in an eigenstate of $\sigma_x$
we now describe a protocol to prepare symmetric and antisymmetric superpositions of squeezed magnon vacua.~Consider an initial product state consisting of the magnon vacuum and an equal superposition of the $\bar{\sigma}_x$ eigenstates of the qubit, i.e., $\ket{\psi(0)} = \ket{0}\otimes (\ket{+_x}+\ket{-_x})/\sqrt{2}$, where $\ket{\pm_x}$ denote the eigenstates of $\bar{\sigma}_x$ with eigenvalues $\pm 1$, respectively.
Under the Hamiltonian $H_{\rm cs}^{(I)}$, the qubit and magnon become entangled, and the joint state at time $t$ is
\begin{align}
	\ket{\psi(t)} &= \frac{1}{\sqrt{2}}\left[S(\xi)\ket{0}\otimes\ket{+x} + S(-\xi)\ket{0}\otimes\ket{-x} \right],
%	&= \frac{1}{2}\bigl[\ket{\xi} + \ket{-\xi}\bigr]\otimes\ket{g} + \frac{1}{2}\bigl[\ket{\xi} - \ket{-\xi}\bigr]\otimes\ket{e}\nonumber,
\end{align}
where $\ket{\xi} = S(\xi)\ket{0}$ denotes a squeezed magnon vacuum state.~A projective measurement of the qubit in the $\sigma_z$ eigenbasis $\{\ket{g},\ket{e}\}$ then collapses the magnon state into either the symmetric or antisymmetric superposition of two orthogonally squeezed vacuum states.~Specifically, conditioned on the measurement outcome $\ket{g}$ or $\ket{e}$, the magnon is projected onto
\begin{align}
	\ket{\psi_\pm} = \frac{1}{\sqrt{\mathcal{N}\pm}}\left[\ket{\xi} \pm \ket{-\xi}\right],
\end{align}
where $\mathcal{N}_\pm = 2[1 \pm \cosh^{1/2}(2r)]$ are normalization constants, and these states correspond to even and odd superpositions of orthogonally squeezed magnon vacua, respectively.

To characterize the nonclassical features of the generated magnon states, we evaluate the Wigner function $W(\alpha)$, a quasiprobability distribution that provides a complete phase-space representation of the density operator.~The Wigner function is defined as
\begin{equation}
	W(\alpha) = \frac{2}{\pi}\mathrm{Tr} [(-1)^{m^\dagger m}D^{\dagger}(\alpha)\rho D(\alpha) ],
\end{equation}
where $D(\alpha)=\exp(\alpha m^\dagger-\alpha^{*}m)$ is the displacement operator with complex amplitude $\alpha$, and $(-1)^{m^\dagger m}$ is the magnon parity operator.~The non-Gaussian character of the superposition states is directly revealed in their phase-space structure.~As shown in Fig.~\ref{fig:fig5}, the Wigner functions of the symmetric and antisymmetric states, evaluated at $t = 29~\mathrm{ns}$, exhibit pronounced interference fringes and negative regions arising from the coherent superposition of orthogonally squeezed components.~These phase-space signatures are a clear hallmark of nonclassicality and make the states a useful resource for non-Gaussian quantum computation~\cite{albarelli2018nongauss}.~The statistical and interference properties of general superpositions of squeezed states with different phases have been examined previously~\cite{sanders1989super}.~More recently, such states have been proposed as a resource for generating heralded single photons~\cite{azuma2024single}.

\begin{figure}[!t]
	\centering
	\includegraphics[width=0.99\linewidth]{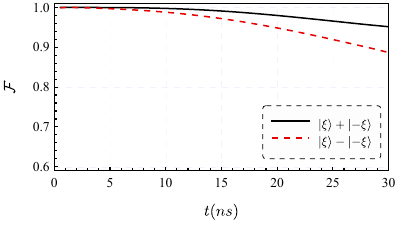}
	\caption{Fidelity of the symmetric ($\ket{\psi_+}$, black solid) and antisymmetric ($\ket{\psi_-}$, red dashed) magnon superposition states, obtained by numerically solving Eqs.~\eqref{eq:me} and~\eqref{eq:me2} and conditioning on a projective measurement in the $\{\ket{g},\ket{e}\}$ basis.}
	\label{fig:fig6}
\end{figure}

To analyze the sensitivity of state preparation to the dissipation in Eq.~\eqref{eq:me2}, we plot in Fig.~\ref{fig:fig6} the fidelity between the symmetric superposition generated without dissipation and that obtained in the presence of dissipation for $\kappa/2\pi = 0.5~\mathrm{MHz}$, $\gamma/2\pi = 3~\mathrm{kHz}$, and $T = 10~\mathrm{mK}$.~Here the fidelity between states with density operators $\rho_1$ and $\rho_2$ is defined as $\mathcal{F} = \text{Tr}[\sqrt{\sqrt{\rho_1}\rho_2\sqrt{\rho_1}}]$.~Figure~\ref{fig:fig6} shows $\mathcal{F}$ as a function of the evolution time $t$: panel (a) corresponds to the symmetric superposition, while panel (b) shows the antisymmetric one.~For the parameters considered here, the fidelities remain close to unity over the relevant time window and exhibit only a mild decay.~The antisymmetric state degrades faster, consistent with its more pronounced interference fringes and hence higher sensitivity to decoherence.

Squeezed superposition states have important applications in quantum error correction.~By constructing symmetric and antisymmetric superpositions of squeezed vacuum states, one can define a logical-qubit encoding analogous to the two-component cat code~\cite{cochrane1999macroscopically}.~This squeezed-superposition encoding exhibits twofold symmetry, with logical states
$|0\rangle_L = [S(\xi)+S(-\xi)]\ket{0}/\sqrt{\mathcal{N}_+}$ and 
$|1\rangle_L = [S(\xi)-S(-\xi)]\ket{0}/\sqrt{\mathcal{N}_-}$.~
Here,
$S(\xi)\ket{0} = \sum_{m=0}^\infty(-1)^m\sqrt{(2m)!}e^{i m \varphi}\tanh^m r/(2^mm!)\ket{2m}/\cosh r$,
so that the logical states contain only terms of the form $|4m\rangle$ and $|4m+2\rangle$ respectively in the Fock basis, thereby exhibiting properties equivalent to a four-component squeezed cat code in phase space.~In the large-squeezing limit, the phase properties of squeezed vacuum states approach those of a superposition of two phase states with opposite phases, forming approximate number-phase codes.~Evaluation using the Knill-Laflamme conditions demonstrates that this encoding can detect single-photon loss and dephasing errors.~Although the Knill-Laflamme conditions are only approximately satisfied in the large-squeezing limit due to the finite separation between logical states in Fock space, increasing the squeezing strength $r$ causes the ratio of expectation values $\langle(m^\dagger m)^p\rangle$ in the two logical states to approach unity, rendering the logical states progressively more indistinguishable and enabling effective protection via quantum error correction~\cite{HopePRR2026}.

\section{\label{sec:level5}Conclusion}

In conclusion, we have introduced and analyzed a scheme for engineering strong magnon squeezing and measurement-conditioned squeezed superposition states in a flux-qubit-YIG hybrid system by exploiting intrinsic longitudinal coupling.~Under a strong microwave drive applied to the qubit, we derived an effective conditional squeezing Hamiltonian that realizes qubit-state-dependent parametric amplification of the Kittel mode.~Numerical simulations incorporating realistic dissipation show that squeezing beyond $8~\mathrm{dB}$ can be achieved in parameter regimes compatible with current experiments and indicate that magnon damping constitutes the dominant limitation to the attainable performance.~By preparing the qubit in a superposition state and post-selecting on projective measurement outcomes, we generate symmetric and antisymmetric superpositions of orthogonally squeezed magnon states.~The resulting phase-space interference provides a direct signature of their non-Gaussian character and underscores their relevance for continuous-variable quantum information tasks.~Moreover, these states naturally implement a logical-qubit encoding with fourfold rotational symmetry, forming approximate number-phase codes capable of detecting single-magnon loss and dephasing errors~\cite{Ayyash2024pra,HopePRR2026}.~Taken together, our results establish the flux-qubit-YIG platform as a practical route to nonclassical magnon state engineering and motivate further investigations into state verification, feedback-based stabilization, and extensions to multimode magnonic architectures for error-corrected quantum information processing.

\begin{acknowledgments}

  The authors thank Wei Wiong for helpful discussion. This work was Supported by the National Natural Science Foundation of China (Grant No.~12247101), the Fundamental Research Funds for the Central Universities (Grant No.~lzujbky-2024-jdzx06) and Youth Science and Technology Fund of Gansu Province (Grant No.~24JRRA997).

\end{acknowledgments}

\section*{Data Availability}

The data that support the findings of this article are openly available \cite{Liugithub2026}.

\end{document}